\documentclass[11pt]{article}
\usepackage{latexsym}
\usepackage{graphicx}
\usepackage{epsfig}
\newcommand \be {\begin{equation}}
\newcommand \bea {\begin{eqnarray}}
\newcommand \ee {\end{equation}}
\newcommand \eea {\end{eqnarray}}

\newcommand{\bit}{\begin{itemize}}
\newcommand{\eit}{\end{itemize}}

\begin{document}
\rightline{MIT-CTP-3467}
\rightline{hep-th/0401124}
\begin{center}
{\bf Topological Gravity, Kaluza-Klein Reduction, and the
Kink}\footnote{Talk given at the {\sl X Marcel Grossmann Meeting},
Rio de Janeiro, Brasil. Based on \cite{gijp}.}

\smallskip

Alfredo Iorio

{\sl Center for Theoretical Physics, Massachusetts Institute of Technology}\\
{\sl 77, Massachusetts Avenue - Cambridge MA 02139 USA} \\

\begin{abstract}
\noindent Kaluza-Klein reduction of the 3d gravitational
Chern-Simons term leads to a 2d theory that supports a symmetry
breaking solution and an associated kink interpolating between AdS
and dS geometries.
\end{abstract}
\end{center}

\section{Chern-Simons Gravity}

Chern-Simons (CS) gauge theories, either pure or as
modifications of Yang-Mills theories, have been studied from a
variety of perspectives, gravity included. Nonetheless, some
aspects of the impact on gravity deserve further investigation.

\noindent The probably best known arena for CS gravity is in 3
dimensions\cite{carlip}
\begin{equation}\label{first}
  {\rm GR}(G_{\mu \nu}) = \int d^3 x \sqrt{G} R \;,
\end{equation}
with $G = {\rm det} G_{\mu \nu}$, $\mu, \nu = 0,1,2$, $\eta_{A B}
= {\rm diag} (+1, -1, ...)$, and $R$ is the scalar curvature. In such a
theory there are no local degrees of freedom, as in $n$ ($\geq 3$)
dimensions they are $(n-3)n/2$. The curvature is fully determined
by its ``Ricci part'', the Weyl tensor vanishing identically:
$R^{\mu \nu}_{\; \; \; \rho \sigma} = - \; \epsilon^{\mu \nu \lambda}
  \epsilon_{\rho \sigma \kappa} E^\kappa_\lambda$, where
$E_{\mu \nu} = R_{\mu \nu} - 1/2 G_{\mu \nu} R$. All solutions of
the Einstein equations are conformally flat, hence in source-free
regions ($T_{\mu \nu}^{\rm matter} = 0$) spacetime is flat (for
$\Lambda =0$). 

\noindent In the early 80's it was noticed that this model
has several interesting properties\cite{djt}, especially when a
CS term is added to ${\rm GR}(G_{\mu \nu})$
\begin{equation}\label{third}
{\rm CS} (\Gamma) = \frac{1}{4 \pi^2} \int d^3 x \epsilon^{\mu \nu
\lambda} \left( \frac{1}{2} \Gamma^\rho_{\mu \sigma} \partial_\nu
\Gamma^\sigma_{\lambda \rho} + \frac{1}{3} \Gamma^\rho_{\mu
\sigma} \Gamma^\sigma_{\nu \tau} \Gamma^\tau_{\lambda \rho}
\right) \;.
\end{equation}
This leads to a topologically massive scalar graviton theory,
descending from the theory ${\rm GR}(G_{\mu \nu}) + 1/\mu \; {\rm
CS} (\Gamma)$. 

\noindent In the first order formulation of Vielbein
($E^A_\mu$) and spin connections ($\Omega^{A B}_\mu$) the gauge
theoretical nature of 3d gravity becomes more transparent.
Treating $E$ and $\Omega$ as independent variables, it was proved
the equivalence\cite{witten} ${\rm GR}(E, \Omega) \Leftrightarrow {\rm CS} (E,
\Omega)$. This only happens in 3 dimensions, where a
non-degenerate, ISO(2,1)-invariant, bilinear form can be defined
in terms of ${\bf P}_A$ and ${\bf J}_A$, generators of
translations and rotations, respectively. The Dreibein is the
gauge field associated to ${\bf P}_A$, while the spin connection
is the gauge field associated to ${\bf J}_A$. A different
perspective takes $\Omega (E)$, obtained imposing torsionlessness
rather than having it has an Euler-Lagrange (EL) equation. On this
see, for instance, \cite{hornewitten}.

\section {Kaluza-Klein Reduction}

\noindent We shall concentrate on CS($\Gamma$) alone and dimensionally reduce 
it in a Kaluza-Klein (KK) setting\cite{gijp}. We have
\begin{equation} \delta {\rm CS} (\Gamma) =
- \frac{1}{8 \pi^2} \int d^3 x \; \sqrt{G} C^{\mu \nu} \delta
G_{\mu \nu} \label{fourth} \;,
\end{equation}
where the Cotton tensor $C^{\mu \nu}$, defined by
\begin{equation}
C^{\mu \nu} = - \frac{1}{2} \frac{1}{\sqrt{G}} (\epsilon^{\rho \sigma
\mu} D_\rho R^\nu_{\sigma} + \mu \leftrightarrow \nu ) \;,
\label{fifth}
\end{equation}
in 3d plays the role of the Weyl tensor. Notice that $C^{\mu \nu}$
exists in any dimension, but only in 3d its vanishing ensures a
necessary and sufficient condition for conformal flatness
\cite{heyl}. Thus, this theory, entirely governed by CS($\Gamma$),
has EL equations $C^{\mu \nu} = 0$, whose solutions are all
conformally flat. 

\noindent Our gauge theory/gravity theory dictionary
consists in reading the Christoffel connection
$\Gamma^\lambda_{\mu \nu}$ as having space-time index $\mu$, and
matrix indices ($\lambda, \nu$): $\left( A_\mu
\right)^\lambda_{\;\;\nu}$. Thus, under general coordinate
transformations
\begin{equation} \label{sixth}
\left( A_\mu (x) \right)^\lambda_{\;\; \nu}  \to  \left(
\tilde{A}_\mu (\tilde{x}) \right)^\lambda_{\;\; \nu}
  = \frac{\partial {x}^\sigma}{\partial
  \tilde{x}^\mu} \left( U ^{-1} A_\sigma (x) U + U^{-1} \frac{\partial }{\partial
  {x}^\sigma} U \right)^\lambda_{\;\;\; \nu}  \;,
\end{equation}
with gauge function $U = \partial x /
\partial \tilde{x}$. 

\noindent Moreover, $R^\rho_{\; \sigma \mu \nu} = (\partial_\nu
\Gamma^\rho_{\mu \sigma}+ \Gamma^\rho_{\nu \tau} \Gamma^\tau_{\mu
\sigma} -  \mu \leftrightarrow \nu )$, coincides with $\left(
F_{\nu \mu} \right)^\rho_{\;\; \sigma} = \left(\partial_\nu A_\mu
+ A_\nu A_\mu - \mu \leftrightarrow \nu \right)^\rho_{\;\;
\sigma}$. Denoting the Dreibein by $E^A_{\;\; \mu}$ (and its
inverse by $E^\mu_{\;\; A}$), the metricity condition defines the
spin connection $\left( \Omega_\mu \right)^A_{\;\; B}$: $D_\mu
E^A_{\;\; \nu} =
\partial_\mu E^A_{\;\; \nu} - \Gamma^\lambda_{\mu \nu} E^A_{\;\;
\lambda} = - \left( \Omega_\mu \right)^A_{\;\; B} E^B_{\;\; \nu}$
or $\left( A_\mu \right)^\lambda_{\;\; \nu} = E_{\;\; A}^\lambda
\left( \Omega_\mu \right)^A_{\;\; B} E^B_{\;\; \nu} + E_{\;\;
A}^\lambda \partial_\mu E^A_{\;\; \nu}$. Thus $A_\mu$ is the gauge
transform of the spin connection $\Omega_\mu$ with the gauge
function $U^B_{\;\; \nu} = E^B_{\;\; \nu}$. It then follows that
$R^A_{\;\; B \mu \nu} = \left( \partial_\nu \Omega_\mu +
\Omega_\nu \Omega_\mu
  - \mu \leftrightarrow \nu \right)^A_{\;\; B}$,
and $R^\rho_{\; \sigma \mu \nu} = E^\rho_{\;\; A} R^A_{\;\; B \mu
\nu} E^B_{\;\; \sigma}$, $R_{\; \sigma \nu} = E^\mu_{\;\; A}
R^A_{\;\; B \mu \nu} E^B_{\;\; \sigma}$, $R = E^\mu_{\;\; A} R^{A
B}_{\;\;\mu \nu} E_B^{\;\; \nu}$. 

\noindent We can write 
\begin{eqnarray}
{\rm CS} (\Gamma) &=& \frac{1}{4 \pi^2} \int d^3 x \epsilon^{\mu \nu
\lambda} \left[ {\rm tr} \left( \frac{1}{2} \Omega_{\mu}
\partial_\nu \Omega_{\lambda} + \frac{1}{3}
\Omega_{\mu} \Omega_{\nu} \Omega_{\lambda} \right) - \frac{1}{6}
{\rm tr} \left(V_{\mu} V_{\nu} V_{\lambda} \right) \right] \\
& = & {\rm CS} (\Omega) + W (E) \;,
\end{eqnarray}
where $(V_\mu)^{\sigma}_{\;\; \rho} = E^\sigma_{\;\; A}
\partial_\mu E^A_{\;\; \rho}$. The last term $W(E)$ is the winding
number of the Dreibein, whose variation is a surface term.

\noindent Therefore variations of CS($\Gamma$) coincide with those of
CS($\Omega$) and $W(E)$ does not contribute to the EL equations.
In standard gauge theories $W(U)$ can give rise to
quantization of the CS term, while this is not the case for the
gravity term\cite{percacci}. We shall not consider $W(E)$ any
further. In 3d $\Omega_{\mu \;, A B} = \epsilon_{A B C} \Omega^C_{\;\;
\mu}$, thus
\begin{equation} \label{nineth}
{\rm CS} (\Omega) = - \frac{1}{4 \pi^2} \int d^3 x \epsilon^{\mu \nu
\lambda} \eta_{A B} \Omega^A_{\; \; \mu} \partial_\nu \Omega^B_{\;
\; \lambda} + \frac{1}{2 \pi^2} \int d^3 x \; {\rm det} \;
\Omega^A_{\; \; \mu} \;.
\end{equation}

\noindent With the KK {\it Ansatz} the 3d metric tensor reads (see
for instance \cite{lor})
\begin{equation} \label{tenth}
  G_{\mu \nu} = \phi \left( \begin{array}{ccc}
    g_{\alpha \beta} - a_\alpha a_\beta &  & - a_\alpha \\
    - a_\beta &  & - 1 \
  \end{array}\right) \;,
\end{equation}
where all quantities are independent of $y$, and, transforming
$G_{\mu \nu}$ under $\delta x^\mu = - \xi^\mu (t,x)$, it is seen
that $g_{\alpha \beta}$ is the 2d metric tensor, $a_\alpha$ is a
2d gauge vector, ($\alpha, \beta = 0,1$), and $\phi$ is a scalar.
As the CS term is conformally invariant we set $\phi = 1$. 

\noindent The reduced Dreibein $E^A_{\;\;\ \mu}$ and spin connection
$\Omega^A_{\;\;\ \mu}$ read
\begin{equation}
  E^a_{\;\; \alpha} = e^a_{\;\; \alpha} \;, \; E^2_{\;\; \alpha} = a_\alpha \;, \;
  E^a_{\;\; 2} = 0 \;, \; E^2_{\;\; 2} = 1 \;,
\end{equation}
\begin{equation}
   \Omega^a_{\;\; \alpha} =  \frac{1}{2} e^a_{\;\; \alpha} f
 \;, \; \Omega^2_{\;\; \alpha} = - \omega_\alpha - \frac{1}{2} f a_\alpha
\;, \; \Omega^a_{\;\; 2} = 0 \;, \; \Omega^2_{\;\; 2} = -
\frac{1}{2} f \;,
\end{equation}
respectively. Here $a,b,... = 0,1$, $e^a_{\; \alpha}$ is the
Zweibein, the 2d spin connection is $\omega_{\alpha , \; a b} =
\epsilon_{a b} \; \omega_\alpha$, $f_{\alpha \beta} =  \sqrt{- g} \;
\epsilon_{\alpha \beta} f$, where $f_{\alpha \beta} =
\partial_{[\alpha} a_{\beta]}$. 

\noindent With these, the dimensionally
reduced gravitational CS term is eventually obtained
\begin{equation}
  {\rm CS}  = - \frac{1}{8 \pi^2} \int d^2 x \sqrt{-g} (f r + f^3)
  \;,
\end{equation}
where $g = {\rm det} g_{\alpha \beta}$, and $r$ is the 2d scalar
curvature. 

\noindent The 3d scalar curvature $R$ with the KK {\it Ansatz}
(and $\phi = 1$) reduces to 
\begin{equation}
R = r + \frac{1}{2} f^2 \;.
\end{equation}

\section{The Kink}

\noindent Variation of reduced action produces
\begin{equation}
  \delta {\rm CS} = \frac{1}{4 \pi^2} \int d^2 x \sqrt{- g}
  (- j^\alpha \delta a_\alpha + \frac{1}{2} T_{\alpha \beta} \delta g^{\alpha
  \beta})\;,
\end{equation}
where $j^\alpha =  - ( 1 / 2 \sqrt{-g} ) \epsilon^{\alpha \beta}
\partial_\beta (r + 3 f^2)$, and $T_{\alpha \beta} = g_{\alpha \beta} (D^2 f - f^3 - \frac{1}{2}
r f) - D_\alpha D_\beta f$. As a consequence of gauge and 2d
diffeo invariance $D_\alpha j^\alpha = 0$, and $D^\beta T_{\alpha
\beta} = 0$, respectively. The components of the dimensionally
reduced Cotton tensor are $C^{\alpha \beta} = T^{\alpha \beta}$,
$C^{\alpha 2} = - j^\alpha - T^{\alpha \beta} a_\beta$, and
$C^{22} = g_{\alpha \beta } T^{\alpha \beta} + a_\alpha T^{\alpha
\beta} a_\beta + 2 j^\alpha a_\alpha$. Thus the EL equations are
\begin{equation}
\epsilon^{\alpha \beta}
\partial_\beta (r + 3 f^2) = 0 \;, \quad \quad g_{\alpha \beta} (D^2 f - f^3 - \frac{1}{2} r f) -
D_\alpha D_\beta f = 0 \;.
\end{equation}
The first is solved by $r + 3 f^2 = {\rm constant} = c$.
Eliminating $r$ in the second equation, and decomposing into the
trace and trace-free parts lead to
\begin{eqnarray}
0 & = & D^2 f - c f + f^3 \;, \\
0 & = & D_\alpha D_\beta f - \frac{1}{2} g_{\alpha \beta} D^2 f
\;.
\end{eqnarray}
The equations are invariant against $f \leftrightarrow - f$. A
solution that respects this {\it symmetry} is: $f = 0$, $r = c$,
with $R=c$. However, there is also a symmetry breaking solution $f
= \pm \sqrt{c}$, $r = - 2 c$, ($c > 0$), with $R = - ( 3 / 2) c$.
The latter is maximally symmetric, the former is not. 

\noindent When the symmetry breaking solution is present, there also is a kink
solution
\begin{equation}
  f = \sqrt{c} \tanh \frac{\sqrt c}{2} x \;,
\end{equation}
interpolating between $f = \pm \sqrt{c}$, and giving rise to 
$ r = - 2 c + 3 c / (\cosh^2 \frac{\sqrt c}{2} x) $,
with 3d scalar curvature $ R = - 3 c / 2 + 5 c / 
(2 \cosh^2 \frac{\sqrt c}{2} x)$.

\noindent The global properties of the reduced theory above described have been now
extensively studied\cite{grumetal}.

\section{Overview}

\noindent There are various dimensional reductions/enhancements
one can perform on the CS gravity term: $3d \rightarrow 2d$ [this
work, \cite{gijp}]; $4d \leftarrow 3d$ \cite{jp}; $(2N+1)d
\rightarrow 2Nd$ in particular $5d \rightarrow 4d$
\cite{grumilleriorio}.

\end{document}